\documentclass[runningheads]{llncs}
\usepackage{graphicx}
\usepackage{listings}
\usepackage{url}
\usepackage{hyperref}
\usepackage{todonotes}
\usepackage{menukeys}

\newcommand{\var}[1]

\renewcommand{\int}{\ensuremath{\mathit{int}}}

\newcommand{\figlabel}[1]{\label{figure:#1}}
\newcommand{\nakedfigref}[1]{\ref{figure:#1}}
\newcommand{\figref}[1]{Figure~\nakedfigref{#1}}

\newcommand{\seclabel}[1]{\label{sec:#1}}
\newcommand{\nakedsecref}[1]{\ref{sec:#1}}
\newcommand{\secref}[1]{Section~\nakedsecref{#1}}

\graphicspath{{./pictures/}}

\begin{document}
\title{Invited Paper: Failure is (literally) an Option:
Atomic Commitment vs Optionality in Decentralized Finance
\thanks{Supported by NSF grant 1917990.}}
\titlerunning{Failure is an option}
%
\author{Daniel Engel\inst{1} \and
Maurice Herlihy\inst{1}\orcidID{0000-0002-3059-8926} \and
Yingjie Xue\inst{1}}
\authorrunning{D. Engel et al.}
%
\institute{Brown University Computer Science Dept, Providence RI 02912, USA }
\maketitle              

\begin{abstract}
Many aspects of blockchain-based decentralized finance
can be understood as an extension of classical distributed computing.
In this paper,
we trace the evolution of two interrelated notions: failure and fault-tolerance.
In classical distributed computing,
a failure to complete a multi-party protocol is typically attributed to hardware malfunctions.
A fault-tolerant protocol is one that responds to such failures
by rolling the system back to an earlier consistent state.
In the presence of Byzantine failures,
a failure may be the result of an attack,
and a fault-tolerant protocol is one that ensures that attackers will be punished
and victims compensated.
In modern decentralized finance however,
failure to complete a protocol can be considered a legitimate option, not a transgression.
A fault-tolerant protocol is one that ensures that the party offering
the option cannot renege,
and the party purchasing the option provides
fair compensation (in the form of a fee) to the offering party.
We sketch the evolution of such protocols,
starting with two-phase commit, and finishing with timed hashlocked smart contracts.
\end{abstract}

\section{Introduction}
\seclabel{intro}
Decentralized finance (DeFi) is on the rise:
between June and October 2020,
the value of assets managed by DeFi protocols
increased from \$1 billion to \$7.7 billion~\cite{defipulse}.
This paper is an informal tutorial,
explaining certain basic problems in DeFi
as if they were problems in fault-tolerant
distributed computing.
Conversely, many core problems in DeFi represent
interesting and important extensions of distributed computing problems.
The goal of this paper is to encourage distributed computing
researchers to consider the kinds of problems and models
that arise in DeFi,
and conversely, to encourage DeFi researchers
to benefit from the rich history of distributed computing techniques
and algorithms.

The contribution of this work is simply to illustrate these
claims through an extended example,
presented with the hope of provoking others to take up research in this area.
We explore how one core problem of distributed computing has evolved over time,
gradually turning into a superficially distinct core problem of finance.
We consider the problem of \emph{atomic commitment}:
how can we install updates at multiple databases or ledgers
in such a way that guarantees that if all goes well, all updates are installed,
but if something goes wrong, all updates are discarded.
The classical challenge is, of course, tolerating failures:
databases can crash or communication can be lost or delayed.

This is one of the oldest problems in distributed computing,
and not surprisingly, it is central to key problems in DeFi.
We explore two aspects of this problem.
First, we explore the technical solutions,
where DeFi has tended to borrow, whether consciously or not,
from prior solutions in distributed computing.
Second, we explore underlying conceptual frameworks,
where DeFi extends the notion of fault-tolerance
well beyond the classical models of distributed computing.
We hope that our examples illustrate how each field
can learn from the other.

In \secref{twophase},
we review the well-known \emph{two-phase commit protocol}~\cite{BernsteinHZ1986},
a classical technique for making atomic updates to independently-failing
databases in a distributed system.
In \secref{atomicity},
we consider the \emph{cross-chain atomic swap} problem,
where mutually-suspicious parties exchange assets
atomically across distinct blockchains.
The simplest atomic swap protocols are based on
\emph{hashed timelocked contracts}~\cite{Herlihy2018,tiersnolan} (HTLCs).
Technically, HTLC protocols closely resemble classical two-phase commit.
The principal difference between the two protocols is in their underlying conceptual frameworks.
In two-phase commit, a failure is typically an operational malfunction at a node or a network,
while in atomic swap, a failure could also be a malicious action chosen by an adversarial party.

This distinction becomes more pronounced in \secref{optionality}.
In both two-phase commit and atomic cross-chain swap protocols,
fault tolerance means that if one party falls silent in the middle,
the other parties are eventually made whole:
database replicas are eventually restored,
and escrowed assets are eventually refunded.
For distributed computing's two-phase commit, the story ends there,
but for DeFi's atomic swap protocol,
the story has just begun.
In finance, the ability to abandon or to complete an in-progress swap
is called an \emph{option}, and options themselves have value.
Any party who abandons an atomic swap should compensate the other parties
by paying a small fee called a \emph{premium}.
Treating failures as compensated options is alien to classical distributed computing models,
where all parties implicitly are on the same team,
but it opens up a range of new research challenges for distributed computing.
Incorporating premiums into atomic swaps turns out to be a challenging technical problem
\cite{XueH2021},
effectively requiring nesting one atomic commitment protocol within another.

\secref{transfer} takes the notion of optionality to the next level.
What if one party could sell such an option to another?
Alice, who has paid for the option to complete or cancel a swap,
should be able to transfer that option to Bob for a fee.
Alice would relinquish her power over the swap's outcome,
and Bob would assume all of Alice's power,
including the power to complete or cancel the swap,
and the right to be compensated if another party cancels the swap.
This problem is also technically challenging,
as it requires embedding yet another atomic commitment mechanism
within other nested atomic commitment mechanisms.

While we advocate thinking about DeFi mechanisms as if they were distributed computing problems,
we also advocate DeFi as a rich source of new problems and models for
mainstream distributed computing.
Originally,
abandoning an atomic commitment protocol was considered a simple operational failure,
and the meaning of fault-tolerance was simply to restore integrity and availability.
When the parties become autonomous and potentially adversarial, however,
failures can become deliberate choices,
and the meaning of fault-tolerance must be extended to provide financial
compensation to any victims of other parties' choices.
Once failures become options (in the financial sense),
then those options themselves become assets to be traded.

The questions raised here are not really about blockchains, as blockchains.
Instead,
they are really about the scientific and engineering problems of safely
transferring value among autonomous parties.
This problem will remain of enduring importance to society,
independently of whether
particular blockchain technologies bloom or fade,
whether certain asset bubbles expand or pop,
or whether regulatory agencies do or do not intervene to protect gullible investors.
We believe the  fault-tolerant distributed computing community has
much to offer on these fundamental problems,
and we encourage the community to get involved.

\section{Model}
\seclabel{model}
A \emph{blockchain} is a tamper-proof distributed ledger or database that
tracks ownership of \emph{assets} by \emph{parties}.
(Our discussion is mostly independent of which blockchain technology is used.)
A party can be a person, an organization, or even a contract (see below).
An asset can be a cryptocurrency, a token, an electronic deed to property, and so on.
There are multiple blockchains managing different kinds of assets.
We focus here on applications where mutually-untrusting parties trade assets among
themselves,
for example by swaps, loans, auctions, markets, and so on.

A \emph{contract} is a blockchain-resident program initialized and called by the parties.
A party can publish a new contract on a blockchain,
or call a function exported by an existing contract.
Contract code and contract state are public,
so a party calling a contract knows what code will be executed.
Contract code must be deterministic because
contracts are typically re-executed multiple times by
mutually-suspicious parties.

Multiple parties agree on a common \emph{protocol} to execute a series of transfers,
an agreement that can be monitored, but not enforced.
Instead of distinguishing between faulty and non-faulty parties,
as in classical distributed computing,
we distinguish only between \emph{compliant} parties
who follow the agreed-upon protocol,
and \emph{deviating} parties who do not.
We make no assumptions about the number of deviating parties.

We assume a \emph{synchronous} execution model
where there is a known upper bound $\Delta$ on the propagation time
for one party's change to the blockchain state to be noticed by the
other parties.
Specifically, blockchains generate new blocks at a steady rate,
and valid transactions sent to the blockchain will be included in a block
and visible to participants within $\Delta$. 
Our example protocols use $\Delta$ as the basis for timeouts:
it is typically chosen conservatively.

We make standard cryptographic assumptions.
Each party has a public key and a private key,
and any party's public key is known to all.
Messages are signed so they cannot be forged,
and they include single-use labels (``nonces'')
so they cannot be replayed.

\section{Classical Two-Phase Commit Protocol}
\seclabel{twophase}
Imagine we have a distributed database with a number of replicas.
These replicas might be identical,
or they may hold different portions of the database (so-called shards).
For simplicity, assume Alice's node holds one replica,
and Bob's node holds another.
A node may \emph{crash} (cease operation),
and later \emph{recover} (resume operation).
Node memory is divided into \emph{volatile} memory lost on a crash,
and \emph{stable} memory that survives crashes.

A \emph{transaction} is a sequence of steps that modifies both replicas.
As a transaction executes, Alice and Bob accumulate a list of tentative changes.
If the transaction \emph{commits}, those changes take effect,
and if the transaction \emph{aborts}, they are discarded.

The \emph{two-phase commit protocol}~\cite{BernsteinHZ1986}
is a classical technique for ensuring \emph{atomicity}:
if a transaction makes tentative changes at both Alice's node and Bob's node,
then the transaction either commits at both nodes or aborts at both.

Here is the simplest form of this protocol.
One node, say Carol, is chosen as the \emph{coordinator}.
\begin{enumerate}
\item \emph{Prepare phase}
  \begin{itemize}
  \item
    Carol, the coordinator, instructs Alice and Bob to record their tentative
    changes in stable storage, so they will not be lost in a crash.
  \item
    If Alice is able to write her changes to stable storage,
    she sends Carol a \emph{yes} vote.
    At this point, some or all of the database becomes inaccessible
    pending the outcome of the transaction.
    If for any reason, Alice cannot save her changes,
    she sends Carol a \emph{no} vote.
    Bob does the same.
  \end{itemize}
\item \emph{Commit phase}
  \begin{itemize}
  \item If Carol receives two \emph{yes} votes, she instructs Alice and Bob to apply
    their tentative changes, committing the transaction.
    If Carol receives a \emph{no} vote, or if either Alice or Bob fails to respond in time,
    she instructs them to discard their tentative changes, aborting the transaction.
    Before Carol sends her decision to Alice and Bob, she records her decision in stable memory, in case she herself crashes.
    
    \item
    Alice follows Carol's instructions.
    If Alice crashes after preparing but before Carol decides,
    Alice must learn the transaction's outcome from Bob or Carol before
    resuming use of her database.
  \end{itemize}
\end{enumerate}
This description is vastly simplified, and omits many practical considerations,
but it serves as a baseline for the more complex DeFi commitment protocols considered in later sections.
The key pattern is that commitment requires that each party agrees to lock up
a set of tentative changes, thereby freezing something of value (here, the database)
until the outcome of the protocol becomes known.

\section{Cross-Chain Atomicity}
\seclabel{atomicity}
Alice has invested in the guilder cryptocurrency,
while Bob has invested in the florin cryptocurrency.
Alice and Bob would both like to diversify:
Alice wants to trade some her guilders for florins,
and Bob wants the opposite trade.
Such an exchange would be almost trivial if both cryptocurrencies
reside on the same chain,
but florins reside on the Florin blockchain,
and guilders on the Guilder blockchain.
Naturally, Alice and Bob do not trust one another,
so we are presented with a more difficult version of last section's
atomic commitment problem:
is there a safe way to guarantee that either both transfers
happen, or neither happens,
\emph{given untrusting participants}.

The two-phase commit protocol is a good start,
but it assumes that all parties are acting in good faith.
Each node reports honestly whether it was able to prepare,
and the coordinator does not lie about the votes it received.
Nevertheless, we can build an atomic cross-chain swap
by ``hardening'' the classical two-phase commit protocol.

We assume each blockchain supports contracts,
and each party can inspect the state of each blockchain.
We make use of a technical gadget called a \emph{hashlock}.
Alice creates a secret value $s$, called the \emph{hashkey}.
She then applies a cryptographic hash function $H$ to $s$,
yielding a (public) \emph{hashlock} $h = H(s)$.
It is effectively impossible to reconstruct $s$ from $h$,
or to find another value $s'$ such that $h = H(s')$.

The notion of \emph{escrow} plays the role of stable storage:
an escrow contract is given custody of Alice's coins,
along with a hashlock $h$ and a timeout.
If $s$ is presented to the contract before the timeout,
then Alice's coins are transferred to Bob,
and if not, those coins are refunded to Alice.
Bob creates a symmetric escrow contract,
only with Alice's hashlock and a different timeout.

Here is the hardened two-phase commit protocol.
\begin{enumerate}
\item \emph{Prepare phase}
  \begin{itemize}
  \item Alice transfers her guilders to her escrow contract with timeout $2\Delta$.
  \item When Bob verifies that Alice's coins have been escrowed,
    he transfers his florins to  his escrow contract with timeout $ \Delta$.
  \end{itemize}
\item \emph{Commit phase}
  \begin{itemize}
  \item When Alice verifies that Bob has put his florins in escrow,
    she sends her secret to the escrow contract on the Florin blockchain,
    unlocking and collecting Bob's florins.
    Alice has now recorded her hashkey on the Florin blockchain.
  \item 
    As soon as Alice's hashkey appears on the Florin blockchain,
    Bob forwards that hashkey to the escrow contract on the Guilder blockchain,
    unlocking and collecting Alice's guilders.
  \end{itemize}
\end{enumerate}
Placing coins in escrow is the analog of writing updates to stable storage
and then voting to commit: each party gives up the ability to back out.
For two-phase commit,
it does not matter which party writes first to stable storage.
For the atomic swap, however,
Alice must escrow first, and Bob second,
because Alice controls the hashkey,
and she could steal Bob's coins if he escrowed first.
The choice of timeouts is critical:
if Bob's timeout were $2\Delta$ instead of $\Delta$,
then Alice could wait until the timeout was about to expire
to claim Bob's florins,
leaving Bob without enough time to claim Alice's guilders.
Atomic swap is less forgiving than two-phase commit:
if Bob falls asleep and fails to claim Alice's guilders
before $2\Delta$ timeout,
then Bob loses the coins on both chains.

A full analysis of this protocol, including failure paths,
is beyond the scope of this paper.
This protocol is called a \emph{hashed timelock contract} protocol.
It was invented by Nolan~\cite{tiersnolan},
generalized to multiple parties~\cite{Herlihy2018},
and used on a number of blockchains
\cite{bitcoinwiki,bip199,decred,barterdex,Catalyst}.

\section{Cross-Chain Atomicity with Optionality}
\seclabel{optionality}
In the previous section,
we argued that one can solve the atomic cross-chain swap
problem by ``hardening'' an existing solution
to the atomic cross-chain commitment problem.
In this section, we argue that the transition from a system
where agents cooperate with one another despite failures,
to a system where agents are potentially adversarial
changes the conceptual framework underlying
common coordination problems.

The HTLC protocol in the last section is safe in the sense
that no compliant party's coins can be stolen.
Each party either completes the swap, or gets its coins back.
Nevertheless,
the HTLC protocol introduces a new problem that
could not have been formulated in the classical distributed computing model:
the \emph{sore loser} attack~\cite{XueH2021}.

Suppose that after Alice escrows her coins,
but before Bob escrows his,
the market shifts,
and Alice's florins lose value with respect to Bob's guilders.
Bob now has the \emph{option} to walk away from the deal,
leaving Alice's coins locked up for a long time,
while Bob is free to use his coins as he pleases.
This problem did not arise in the classical two-phase commit protocol
where all parties' interests were assumed to be aligned.

\paragraph{Premiums}
The problem of optionality is well-understood in the financial world.
If Bob has the option to walk away,
leaving Alice temporarily unable to access her coins, called her \emph{principal},
then Bob should compensate Alice by paying her a small fee, called a \emph{premium}.
There are well-known formulas for computing fair premiums
given asset volatility and escrow duration~\cite{han2019optionality}.
In practice, a 2\% premium is often appropriate.

The problem of adding premiums to atomic swaps is tricky,
because it involves nesting one kind of atomic commitment (the premium deposit)
inside another (the swap).
If the premium is deposited before the principal,
then the principal is protected from sore loser attacks.
But the premium itself is now exposed to a reverse sore loser attack:
what if Alice walks away immediately after Bob escrows his premium?
The way to resolve this ``chicken-and-egg'' problem is to observe
that the value of the premium is much lower than the value of the principal,
and while Alice might not be willing to risk locking up 100 coins,
Bob is probably willing to risk locking up 1 coin.
For very large principals,
Alice and Bob can bootstrap their premiums:
Bob risks 1 coin, Alice escrows 100 coins protected by Bob's 1-coin premium,
Bob escrows 1000 coins protected by Alice's 100-coin premium, and so on.

\paragraph{Two-Party Swap with Premiums}
Here we present a simple two-party swap protocol with premiums,
taken from Xue and Herlihy~\cite{XueH2021}.
Let $p_a$ be the compensation Alice should pay to Bob if Bob is a victim,
and let $p_b$ be the compensation from Bob to Alice.
A contract on the guilder blockchain accepts Alice's principal and Bob’s premium,
and a symmetric contract on the florin blockchain accepts Bob’s escrow and Alice’s premium. 
The timeout for the first step is $\Delta$ from the start of the protocol, and subsequent timeouts increase by $\Delta$.

A straightforward idea is to let Alice deposit premium $p_a$ and Bob $p_b$. 
However, if Alice does not redeem Bob's principal, Bob will not be able to redeem Alice's principal, 
so as a result, Bob pays a premium to Alice, and Alice to Bob.
Therefore, Alice should pay $p_a+p_b$ to Bob in case she does not redeem Bob's principal. 
Here is the protocol, where each step is labeled with its timeout.
See \figref{twopartypremium}.
\begin{enumerate}
    \item[$\Delta$] Alice deposits her premium $p_a+p_b$ on the florin blockchain’s
escrow contract with timelock $t_B = 5 \Delta$.
\item[$2\Delta$] Bob deposits his premium on the guilder blockchain’s
escrow contract with timelock $t_A = 6 \Delta$.
\item[$3\Delta$] Alice escrows her principal on guilder blockchain’s
escrow contract. If she fails to do so, the premium $p_b$ is refunded to Bob. Otherwise, the premium remains in the contract.
\item[$4\Delta$] Bob escrows his principal on florin blockchain’s
escrow contract. If he fails to do so, the premium $p_a+p_b$ is refunded to Alice. Otherwise, the premium remains in the contract.
\item[$5\Delta$] Alice sends a secret $x$ where $H(x)=h$  to redeem Bob's principal.
 If she fails to do so, the premium $p_a+p_b$ in the contract is paid to Bob. If she redeems Bob's principal, the premium is refund to her.
 \item[$6\Delta$] Bob sends a secret $s$ where $H(x)=h$  to redeem Alice's principal.
 If he fails to do so, the premium $p_b$ in the contract is paid to Alice. If he redeems Alice's principal, the premium is refund to him. 
\end{enumerate} 

\begin{figure}
    \centering
    \includegraphics[width =\textwidth]{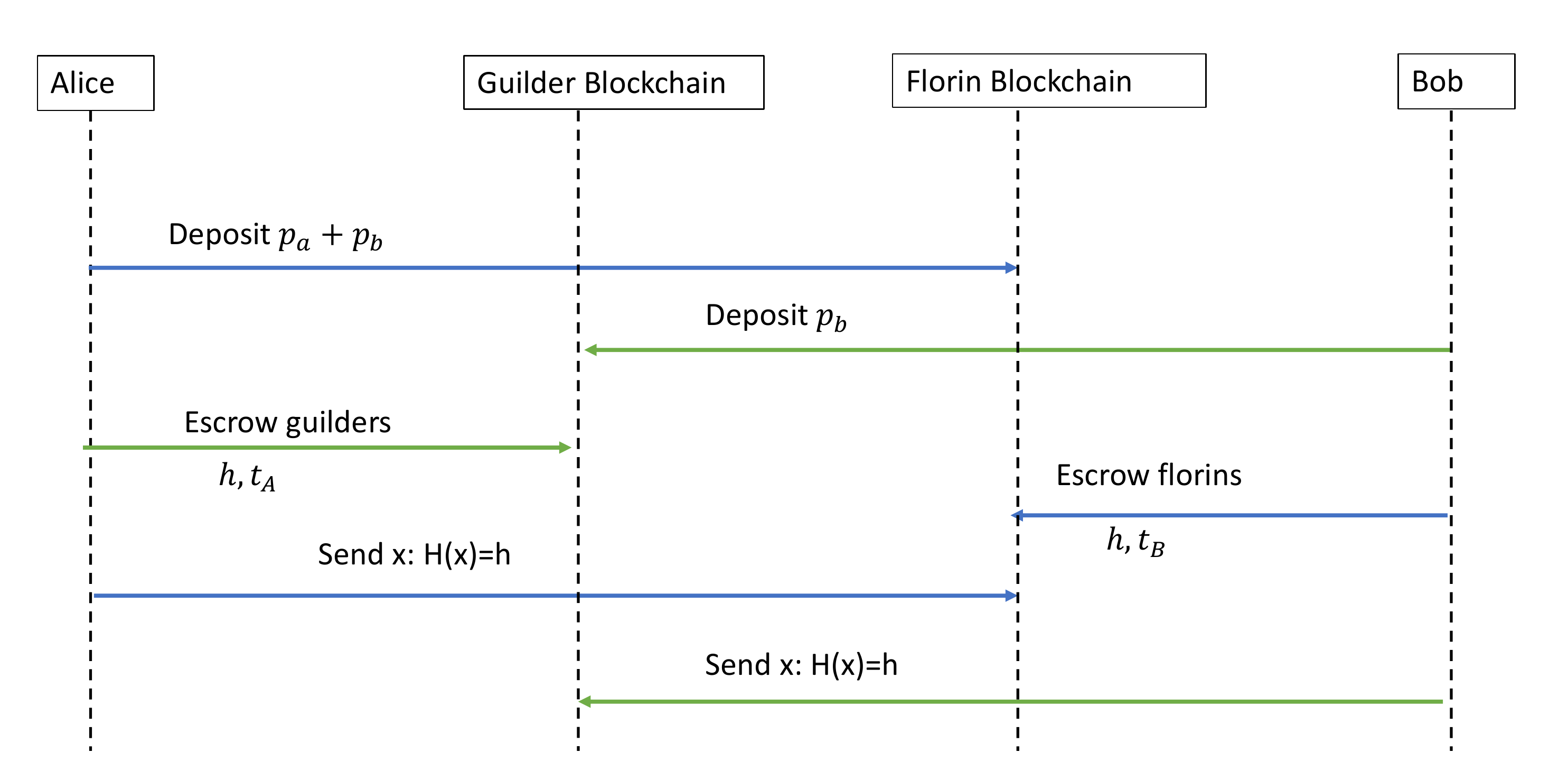}
    \caption{Two-party Swap with Premiums}
    \figlabel{twopartypremium}
\end{figure}

After Alice escrows her principal, if Bob reneges, Alice can get $p_b$ as compensation. If the swap fails after Bob escrows his principal due to Alice, Bob is compensated $p_a$.

The goal of this chapter is to illustrate the progression
from the classical two-phase commit protocol to atomic cross-chain swap,
to atomic cross-chain swap with premiums.
The techniques are recognizably similar:
move the item of value to a safe place,
check that everything is ok, and if so, pull the trigger.
The nature of the problem has shifted in interesting ways:
protocol failures are no longer external events beyond the parties' control,
they have become potentially rational choices
requiring nested atomic commitment mechanisms for protection.
In the following section,
we take optionality to the next level.

\section{Cross-Chain Atomicity with Transferrable Optionality}
\seclabel{transfer}
At this point,
we have shifted the protocol from one where Alice and Bob agree to trade guilders for
florins to one where Alice buys the \emph{option} to make that trade.
If she exercises the option, the swap happens,
and if she declines to do so, she pays Bob a premium for his troubles.

While the option is capable of being exercised, it has value.
It is standard in traditional finance to trade option contracts:
Alice should be able to sell her option with Bob to a third party, Carol.
If Carol buys the option,
she acquires Alice's right to exercise the option before it expires,
and Alice relinquishes all her rights.
As usual, it should be possible for Alice to sell the option to Carol
without placing any compliant party at risk.

Why might Alice want to transfer her option to Carol?
Perhaps Alice has private information suggesting that the relative value of florins to guilders
will change in the near future.
If she does not plan to exercise the option,
then selling it will help pay for her lost premium.

Why might Carol be willing to by an option from Alice?
Perhaps Alice and Carol have asymmetric information:
one thinks florins will increase in value and the other disagrees.
In an illiquid options market,
Carol might have trouble finding a way to buy florins,
so Alice would be a natural counterparty.
In a highly liquid market,
Alice might be willing to offer a discount to dump her option.

Even if Alice and Carol have symmetric information,
they might have different risk tolerances.
Consider the price of a florin expressed in guilders at time $t = 0$.
At $t = 1$, both Alice and Carol believe that with equal probability,
the price will either increase by $dx$ or decrease by $dx$.
If Alice is risk-averse or indifferent,
but Carol is risk-seeking,
then Carol will want to buy that option from Alice,
and Alice will want to sell.

A full protocol for transferable cross-chain options is beyond the scope of this
paper, and appears elsewhere~\cite{futureoption}.
Instead, we present a naive protocol that almost solves the problem,
but the ways in which it falls short are instructive.

\begin{figure}
  \begin{center}
    \begin{tabular}{| c | l |}
      \hline
      Timeout &Action\\ \hline
      &Alice creates $AB$ swap edge with timeout $A: 7\Delta$.\\
      $\Delta$  &Bob creates $BA$ with $A: 6\Delta$.\\
      $2\Delta$ &If Carol does not show up, the protocol proceeds as a normal swap.\\
      &Otherwise Carol creates $CA$ with $C: 9\Delta$.\\
      $3\Delta$ &Alice modifies $AB$ to $A: 7\Delta$ or $C: 8\Delta$.\\
      $4\Delta$ &Bob modifies $BA$ to $A: 6\Delta$ or $C: 7\Delta$.\\
      $5\Delta$ &Alice creates $AC$ swap edge with $C: 7\Delta$.\\
      $6\Delta$ &Carol reveals $C$ on both $BA,AC$.\\
      $7\Delta$ &Alice reveals $C$ on $CA$. \\
      \hline
    \end{tabular}
  \end{center}
  \caption{Partial Protocol for Transferable Option}
  \figlabel{transfer}  
\end{figure}
Here is a na\"ive Transfer Protocol.
For simplicity, we address the easier problem:
how to transfer a position in a 2-party swap (without premiums).
The protocol is shown in \figref{transfer}.
Initially Alice creates a swap with Bob.
If Carol offers to buy the option and Alice agrees,
Alice transfers her position to Carol.
If Alice does not agree,
Alice proceeds with the protocol as normal.
Alice has a secret $A$ and Carol has a secret $C$.

For brevity, we use \emph{edge} $XY$ as shorthand
for a tentative (escrowed) transfer from party $X$ to party $Y$.
The notation $X: k \Delta$ means that the asset on that edge
is transferred if triggered by $X$'s secret before $k \Delta$ time
after the start of the protocol.
``$X: k \Delta$ or $Y: \ell \Delta$'' means the asset
is transferred if either $X$ or $Y$ triggers the transfer by revealing a secret before the respective timeouts.
While this na\"ive protocol conveys the flavor of a full protocol,
there are several reasons it is unsatisfactory.

First,
there is no clear distinction between when Carol buys the swap from Alice,
and when she exercises the swap.
Alice just wants to sell her option and have Carol assume Alice's role immediately.
Here, however, Alice she has to wait for Carol to make up her mind.
Alice should be able to walk away as soon as decides she wants to buy the option.

Whether Carol does nor does not decide to participate,
the ability to sell the option adds $3\Delta$ extra rounds to the original swap protocol.
An ideal protocol would behave like a typical 2-party swap if Carol never participates,
taking the usual $4\Delta$ rounds at most.

Because Alice is entangled in the protocol until Carol decides to exercise it,
Alice has to escrow more than she would otherwise.
That is, she has to escrow assets on $AC$ in addition to the original assets she escrowed on $AB$. 
Alice should only have to escrow what she had in the original swap protocol.

These observations illustrate the challenges of designing transferrable options for
even a simple two-party swap option.
In general,
we would like to be able to transfer more complex, linked options.
For example, in a \emph{cross-chain deal}~\cite{herlihy_cross-chain_2019},
parties can set up a complex network of swaps to be executed atomically,
and a mature DeFi system would allow any party
to sell their position in that network to another party.
Similar challenges arise with types of cross-chain commerce such as bonds, stocks,
and derivatives.

\section{Related Work}
\seclabel{related}
The use of HTLCs for two-party cross-chain swaps is generally attributed to
Nolan~\cite{tiersnolan}.
HTLCs have adapted to several uses~\cite{bitcoinwiki,bip199,decred,barterdex}.
Herlihy~\cite{Herlihy2018} extended HTLCs to support multi-party swaps on directed graphs.

Herlihy \emph{et al.}~\cite{HerlihyLS2021} introduce the notion of
\emph{cross-chain deals}.
They focus on how conventional notions of atomicity are inadequate
for an adversarial environment,
and give protocols using both HTLCs and a central coordinating blockchain.
Zakhary \emph{et al.}~\cite{ZakharyAE2019} propose a 
cross-chain swap protocol for proof-of-work blockchains
using a \emph{witness blockchain} as a central coordinator.

The BAR (byzantine, altruistic and rational)
model~\cite{AiyerACDMP2005,BAR-Primer:2006} supports cooperative services
spanning autonomous administrative domains that are resilient to
Byzantine and rational manipulations.
BAR-tolerant systems assume a bounded number of Byzantine faults,
and as such do not fit our adversarial model,
where any number of parties may be Byzantine, rationally or not.

In finance,
\emph{optionality}~\cite{higham2004introduction} is the notion that
there is value in acquiring the right,
but not the obligation, to invest in something later.
Atomic swap based on HTLCs exposes such optionality to both parties. 
However, multiple researchers~\cite{han2019optionality,arwen2019,liu2018} have observed that both parties are exposed to sore loser attacks where the counterparty reneges at critical points in the protocol.  Robinson~\cite{danrobinson} proposes to reduce vulnerability to sore loser attacks by splitting each swap into a sequence of very small swaps,
an approach that works only for fungible, divisible tokens.

Xue and Herlihy~\cite{XueH2021} show how to incorporate
premiums into multi-party swaps, auctions, and brokered sales.
Prior work was focused exclusively on two-party swaps,
and proposed \emph{asymmetric} protocols,
meaning that only one party pays a premium to the other,
protecting only that side of the swap from a sore loser attack.
These protocols include Han \emph{et al.}~\cite{han2019optionality},
Eizinger \emph{et al.}~\cite{eizinger},
Liu~\cite{liu2018},
the Komodo platform~\cite{komodo},
Eizinger et al. \cite{eizinger},
and the Arwen protocols~\cite{arwen2019}.

Xu \emph{et al.} \cite{xu2020game} analyze the success rate of cross-chain swaps using HTLCs.
Liu~\cite{liu2018} proposed an atomic swap protocol
that protects both parties from sore loser attacks,
structured so that Alice explicitly purchases an option from Bob,
and her premium is never refunded.
There is no obvious way to extend this protocol to applications other than two-party swaps.
Tefagh \emph{et al.}~\cite{tefaghcapital} propose a similar protocol
based on an options model.

\section{Conclusions}
\seclabel{conclusions}
We have argued elsewhere~\cite{mphcacm} that some early blockchain work
recapitulated ideas and algorithms from distributed computing,
sometimes falling prey to familiar pitfalls~\cite{nyt-dao,wsj-dao}.
Here, we argue that blockchain and DeFi open up new opportunities
for distributed computing research.
In this paper,
we outlined how atomic commitment, a classical distributed computing problem,
lies at the heart of several DeFi challenges.
At the same time,
moving from a hardware failure model to a Byzantine failure model
opens up rich new research possibilities.
Dealing with optionality requires nesting one atomic commitment
mechanism inside another (to support premiums),
and fully embracing optionality requires nesting
yet another atomic commitment mechanism (to support option sale and transfer).
We hope that this paper will help draw the attention
of our community to these intriguing questions.

\bibliographystyle{splncs04}
\bibliography{local,references}

\begin{thebibliography}{10}
\providecommand{\url}[1]{\texttt{#1}}
\providecommand{\urlprefix}{URL }
\providecommand{\doi}[1]{https://doi.org/#1}

\bibitem{AiyerACDMP2005}
Aiyer, A.S., Alvisi, L., Clement, A., Dahlin, M., Martin, J.P., Porth, C.:
  {BAR} fault tolerance for cooperative services. In: Proceedings of the
  twentieth {ACM} symposium on operating systems principles. pp. 45--58. {SOSP}
  '05, ACM, New York, NY, USA (2005). \doi{10.1145/1095810.1095816},
  \url{http://doi.acm.org/10.1145/1095810.1095816}

\bibitem{BernsteinHZ1986}
Bernstein, P.A., Hadzilacos, V., Goodman, N.: Concurrency control and recovery
  in database systems. Addison-Wesley Longman Publishing Co., Inc., Boston, MA,
  USA (1986)

\bibitem{bitcoinwiki}
{bitcoinwiki}: Atomic cross-chain trading,
  \url{https://en.bitcoin.it/wiki/Atomic_cross-chain_trading}

\bibitem{bip199}
Bowe, S., Hopwood, D.: Hashed time-locked contract transactions,
  \url{https://github.com/bitcoin/bips/blob/master/bip-0199.mediawiki}

\bibitem{BAR-Primer:2006}
Clement, A., Li, H., Napper, J., Martin, J.P.M., Alvisi, L., Dahlin, M.: {BAR}
  primer. In: Proceedings of the international conference on dependable systems
  and networks ({DSN}), {DCC} symposium (2008), place: Anchorage, AK

\bibitem{decred}
{DeCred}: Decred cross-chain atomic swapping,
  \url{https://github.com/decred/atomicswap}

\bibitem{eizinger}
Eizinger, T., Fournier, L., Hoenisch, P.: The state of atomic swaps.
  \url{http://diyhpl.us/wiki/transcripts/scalingbitcoin/tokyo-2018/atomic-swaps/}
  (2018)

\bibitem{futureoption}
Engel, D., Herlihy, M., Xue, Y.: Transferrable cross-chain options (2021)

\bibitem{han2019optionality}
Han, R., Lin, H., Yu, J.: On the optionality and fairness of {Atomic} {Swaps}.
  In: Proceedings of the 1st {ACM} {Conference} on {Advances} in {Financial}
  {Technologies}. pp. 62--75. ACM, Zurich Switzerland (Oct 2019).
  \doi{10.1145/3318041.3355460},
  \url{https://dl.acm.org/doi/10.1145/3318041.3355460}

\bibitem{arwen2019}
Heilman, E., Lipmann, S., Goldberg, S.: The arwen trading protocols (Jan 2019),
  \url{https://www.arwen.io/whitepaper.pdf}

\bibitem{Herlihy2018}
Herlihy, M.: Atomic cross-chain swaps. In: Proceedings of the 2018 {ACM}
  symposium on principles of distributed computing. pp. 245--254. {PODC} '18,
  ACM, New York, NY, USA (2018). \doi{10.1145/3212734.3212736},
  \url{http://doi.acm.org/10.1145/3212734.3212736}, number of pages: 10 Place:
  Egham, United Kingdom tex.acmid: 3212736

\bibitem{mphcacm}
Herlihy, M.: Blockchains from a distributed computing perspective. Commun. ACM
  \textbf{62}(2),  78–85 (Jan 2019). \doi{10.1145/3209623},
  \url{https://doi.org/10.1145/3209623}

\bibitem{herlihy_cross-chain_2019}
Herlihy, M.: Cross-chain {Deals} and {Adversarial} {Commerce}. CoRR
  \textbf{abs/1905.09743} (2019), \url{http://arxiv.org/abs/1905.09743}

\bibitem{HerlihyLS2021}
Herlihy, M., Liskov, B., Shrira, L.: Cross-chain {Deals} and {Adversarial}
  {Commerce}. Proceedings of the VLDB Endowment  \textbf{13}(2),  100--113 (Oct
  2019). \doi{10.14778/3364324.3364326}, \url{http://arxiv.org/abs/1905.09743},
  arXiv: 1905.09743

\bibitem{higham2004introduction}
Higham, D.J.: An introduction to financial option valuation: mathematics,
  stochastics and computation. Cambridge Univ. Press, Cambridge, 4. printing
  edn. (2009)

\bibitem{liu2018}
Liu, J.A.: Atomic {Swaptions}: {Cryptocurrency} {Derivatives}. arXiv:1807.08644
  [cs, q-fin]  (Mar 2020), \url{http://arxiv.org/abs/1807.08644}, arXiv:
  1807.08644

\bibitem{tiersnolan}
Nolan, T.: Atomic swaps using cut and choose (Feb 2016),
  \url{https://bitcointalk.org/index.php?topic=1364951}

\bibitem{barterdex}
Organization, T.K.: The {BarterDEX} whitepaper: {A} decentralized, open-source
  cryptocurrency exchange, powered by atomic-swap technology,
  \url{https://supernet.org/en/technology/whitepapers/BarterDEX-Whitepaper-v0.4.pdf}

\bibitem{komodo}
Platform, K.: Advanced blockchain technology, focused on freedom.
  \url{https://docs.komodoplatform.com/basic-docs/start-here/core-technology-discussions/introduction.html\#note-on-changes-since-whitepaper-creation-cr-2019}
  (July,2019)

\bibitem{nyt-dao}
Popper, N.: A venture fund with plenty of virtual capital, but no capitalist.
  New York Times (man 2016),
  \url{https://www.nytimes.com/2016/05/22/business/dealbook/crypto-ether-bitcoin-currency.html}

\bibitem{danrobinson}
Robinson, D.: Htlcs considered harmful.
  \url{http://diyhpl.us/wiki/transcripts/stanford-blockchain-conference/2019/htlcs-considered-harmful/}
  (2019)

\bibitem{tefaghcapital}
Tefagh, M., Bagheri, F., Khajehpour, A., Abdi, M.: Capital-free futures
  arbitrage (October, 2020). \doi{10.13140/RG.2.2.31609.90729/1},
  \url{https://www.researchgate.net/profile/Mojtaba-Tefagh-2/publication/344886866_Capital-free_Futures_Arbitrage/links/5fdc88e3a6fdccdcb8d89ee1/Capital-free-Futures-Arbitrage.pdf}

\bibitem{wsj-dao}
Vigna, P.: Chiefless company rakes in more than \$100 million. Wall Street
  Journal (may 2016),
  \url{https://www.wsj.com/articles/chiefless-company-rakes-in-more-than-100-million-1463399393}

\bibitem{xu2020game}
Xu, J., Ackerer, D., Dubovitskaya, A.: A {Game}-{Theoretic} {Analysis} of
  {Cross}-{Chain} {Atomic} {Swaps} with {HTLCs}. arXiv:2011.11325 [cs]  (Apr
  2021), \url{http://arxiv.org/abs/2011.11325}, arXiv: 2011.11325

\bibitem{XueH2021}
{Yingjie Xue}, {Maurice Herlihy}: Hedging {Against} {Sore} {Loser} {Attacks} in
  {Cross}-{Chain} {Transactions}. In: {ACM} {Symposium} on {Principles} of
  {Distributed} {Computing} (2021)

\bibitem{defipulse}
Young, J.: Defi explosion: Uniswap surpasses coinbase pro in daily volume
  (2020)

\bibitem{ZakharyAE2019}
Zakhary, V., Agrawal, D., El~Abbadi, A.: Atomic commitment across blockchains.
  CoRR  \textbf{abs/1905.02847} (2019), \url{http://arxiv.org/abs/1905.02847},
  arXiv: 1905.02847 tex.bibsource: dblp computer science bibliography,
  https://dblp.org tex.biburl:
  https://dblp.org/rec/bib/journals/corr/abs-1905-02847 tex.timestamp: Mon, 27
  May 2019 13:15:00 +0200

\bibitem{Catalyst}
Zyskind, G., Kisagun, C., FromKnecht, C.: Enigma {Catalyst}: a machine-based
  investing platform and infrastructure for crypto-assets,
  \url{https://www.enigma.co/enigma_catalyst.pdf}

\end{thebibliography}
\end{document}